\documentclass[11pt,epsf,letterpaper]{article}%
\pdfoutput=1
\usepackage{setspace}
\usepackage{color}
\usepackage{amsmath}
\usepackage{amsfonts}
\usepackage{verbatim}
\usepackage{amssymb}
\usepackage{graphicx}
\usepackage{amsmath,amssymb,calc}
\usepackage[bulletsep]{collref}
\usepackage{bbm}
\usepackage{setspace}
\usepackage{color}
\usepackage{amsmath}
\usepackage{amsfonts}
\usepackage{verbatim}
\usepackage{amssymb}
\usepackage{graphicx}
\usepackage{array}
\usepackage{caption}
\usepackage{subcaption}%
\providecommand{\U}[1]{\protect\rule{.1in}{.1in}}
\newcommand\be{\begin{equation}}
\newcommand\ee{\end{equation}}
\newcommand{\bead}{\begin{aligned}}
\newcommand{\eead}{\end{aligned}}

\newcommand{\bea}{\begin{eqnarray}}
\newcommand{\eea}{\end{eqnarray}}

\def\beq{\begin{equation}}
\def\eeq{\end{equation}}
\def\id{\protect{{1 \kern-.28em {\rm l}}}}

\def\unit{\relax{\rm 1\kern-.26em I}}

\begin{document}

\title{Constraining Monopoles by Topology: an Autonomous System}
\author{Fabrizio Canfora$^{1,2}$, Gianni Tallarita$^{1}$\\$^{1}$\textit{Centro de Estudios Cient\'{\i}ficos (CECS), Casilla 1469,
Valdivia, Chile.}\\$^{2}$\textit{Universidad Andres Bello, Av. Republica 440, Santiago, Chile.}\\{\small canfora@cecs.cl,\;\; tallarita@cecs.cl}}
\maketitle

\begin{abstract}
We find both analytical and numerical solutions of $SU(2)$
Yang-Mills with an adjoint Higgs field within both closed and open tubes whose sections
are spherical caps. This geometry admits a smooth limit in which the
space-like metric is flat and, moreover, allows one to use analytical tools which
in the flat case are not available. Some of the analytic configurations, in
the limit of vanishing Higgs coupling, correspond to magnetic monopoles and
dyons living within this tube-shaped domain. However, unlike what happens in the standard case,
analytical solutions can also be found in the case in which the Higgs coupling
is non-vanishing. We further show that the
system admits long-lived breathers.

\end{abstract}

\section{Introduction}

One of the most important discoveries in non-Abelian gauge theories are monopole solutions of the Yang-Mills-Higgs system, with the Higgs
field in the adjoint representation of the gauge group \cite{thooft}
\cite{polyakov} whose stability is ensured by the presence of a topological
charge related to the non-Abelian magnetic field (for two detailed reviews see
\cite{manton} \cite{susy}). After the construction of monopoles,
dyonic solutions carrying both non-Abelian magnetic as well as electric
charges have also been found \cite{Julia:1975ff}. Monopole and dyon solutions have
extensive uses in modern physics, ranging between several research areas which
include many theoretical applications. For example, the fundamental importance of these
non-Abelian solutions in the physics of the early universe is well recognized
(see, for a comprehensive review, \cite{vilenkin}).

For the above reasons, it is extremely useful to have analytic non-Abelian
solutions. In general it is also important to understand how genuinely
three-dimensional topological structures (such as the Yang-Mills-Higgs
hedgehog) react to the presence of spatial boundaries. Indeed, in many of the
most relevant practical applications of Yang-Mills theory the gauge field is
confined to a bounded three-dimensional spatial region. However, in almost all
cases, explicit solutions are available only in the limit in which both the
Higgs coupling $\lambda$ vanishes and the spatial regions where the hedgehogs
live are unbounded in all the three spatial directions \cite{BPS1} \cite{BPS2}.
Even though this case is very important in the context of supersymmetric gauge
theories (see \cite{susy} and references therein), a generalization to bounded regions and non vanishing $\lambda$ is still lacking. The technical tool used in this paper in order to address this issue, is a generalization of the usual hedgehog ansatz
used in \cite{thooft} \cite{polyakov}. A key observation \cite{symmetry1}
\cite{symmetry2} \cite{symmetry3} is that such ansatz realizes spherical
symmetry in a non-trivial way on suitable curved spaces. Namely, even if the gauge potential and the
Higgs field depend explicitly on the angles of the (spherically symmetric)
space-time metric, the corresponding energy-density does not.

Here, using a simple curved background chosen so as to preserve the spherical symmetry of the monopole ansatz, we show that one can construct exact solutions of the Yang-Mills-Higgs
system even in the case in which the Higgs coupling $\lambda$ is non-vanishing. These solutions, unlike those with $\lambda=0$, cannot however be interpreted as magnetic monopoles. 

This paper is organized as follows: in the second section we review the
generalized hedgehog ansatz and introduce the curved system we wish to use,
explaining in detail its physical motivation in the context of this study. In
section 3 we look at some particular magnetic solutions, here we show that
analytic configurations of the gauge-Higgs system can be found corresponding
to magnetic monopoles and breathers even if $\lambda\neq0$. In section 4 we
extend this study to include electric charges in our solution and find dyonic
configurations of the system. Finally, some conclusions will be drawn.

\section{The System}

The action of the $SU(2)$ Yang-Mills-Higgs system in four dimensional
space-time is
\begin{align}
S_{\mathrm{YMH}}  &  =\int d^{4}x\sqrt{-g}\;\mathrm{Tr}\left(  \frac{1}%
{4}F_{\mu\nu}F^{\mu\nu}+\frac{1}{2}D_{\mu}\Psi D^{\mu}\Psi-\frac{\lambda^{2}%
}{8}\left(  \Psi\Psi-v^{2}\right)  ^{2}\right)  \ , \label{skyrmaction}%
\end{align}
where the Planck constant and the speed of light have been set to $1$, and the
coupling constants are $e$ and $\lambda$. Our non-Abelian field strength
reads (we set $e=1/2$ throughout)
\begin{equation}
F_{\mu\nu} = \partial_{\mu}A_{\nu}- \partial_{\nu}A_{\mu}-\frac{1}%
{2}\big[A_{\mu},A_{\nu}\big],
\end{equation}
whilst the covariant derivative is
\begin{equation}
D_{\mu}\Psi= \partial_{\mu}\Psi- \frac{1}{2}[A_{\mu},\Psi],
\end{equation}
with the Higgs field $\Psi$ in the adjoint representation of the gauge group.

We will consider the following metric corresponding to $\mathbb{R}\times S^2$ (or $S^1\times S^2$) in the spatial directions:
\begin{equation}
ds^{2}=-dt^{2}+dr^{2}+R_{0}^{2}(d\theta^{2}+\sin^{2}\theta\ d\phi
^{2})\ ,\ \ 0\leq r\leq L\ ,\label{metric}%
\end{equation}%
\begin{equation}
0\leq\theta\leq\pi\ ,\ 0\leq\phi\leq2\pi\ ,\label{range1}%
\end{equation}
where, as explained in a moment, $L$ is a longitudinal length and
$R_{0}$ is a constant with the dimension of length related to the size of the
transverse sections of the tube. The above choice of metric in eqs. (\ref{metric}) and
(\ref{range1}) is natural for several reasons: first of all, the isometry
group of this metric contains $SO(3)$ as a subgroup and therefore one can
realize the spherical symmetry as usually done for the 't Hooft-Polyakov
monopole living on flat space \cite{symmetry1} \cite{symmetry2}
\cite{symmetry3}. Another important feature of the chosen metric is that,
even though it describes a curved geometry (the non-vanishing components of the Riemann
tensor being proportional to $\left(  R_{0}\right)  ^{-2}$), it admits a smooth flat
space limit $R_{0}\rightarrow\infty$ as all the solutions which will be
considered depend smoothly on $R_{0}$ (thus, in principle, it is possible to
make the curvature effects as small as one wants). However, the present
geometry has the additional benefit of providing an explicit cut-off without
losing the symmetries inherent to the model\footnote{The presence of the
parameters $R_{0}$ and $L$ allow one to take into
account finite-volume effects. Within the present framework, the
Yang-Mills-Higgs system is living in a manifold whose three-dimensional
spatial sections have a finite volume equal to $4\pi LR_{0}^{2}$. In
particular, in the cases in which analytic solutions are available, one can
analyze how the total energy of the system changes with $R_{0}$. We hope to
come back on this issue in a future publication.}. Indeed, one could think of
placing the system whose action is given by eq. (\ref{skyrmaction}) in a box (in flat space), this would
provide an explicit cut-off at the cost of loss of spherical symmetry of the
hedgehog-like configurations. On the other hand, one could use a ``spherical
box" in flat space but, in such a case, it becomes difficult to implement the
physical boundary conditions\footnote{The reason is that in the standard flat
case in polar coordinates the hedgehog profiles of the Higgs and Yang-Mills
fields approach the physical vacua only when $r$ goes to spatial infinity.}.
The present model serves to perform a similar task but retains the original
symmetry characteristic of the finite energy solutions the model possesses in
flat space allowing, at the same time, the use of powerful analytical tools
from the theory of dynamical systems which usually are not available.

One can
\textquotedblleft concretely" realize the metric in eq. (\ref{metric}) as a
cylinder with spherical caps as section. Indeed, if, instead of the full range
of the angular coordinates in eq. (\ref{range1}), one considers the following
restricted range:%
\begin{equation}
0\leq\theta\leq\delta<\frac{\pi}{2}\ ,\ 0\leq\phi\leq2\pi\ .\label{range2}%
\end{equation}
The metric in eqs. (\ref{metric}) and (\ref{range2}) thus represents a tube
(which can be closed or open depending on whether or not one considers the
coordinates $r$ as periodic) whose sections, instead of being disks, are
portions of two-spheres whose size depends on $\delta$. All the configurations
constructed in this paper are exact solutions of the Yang-Mills-Higgs system
both when the range is chosen as in eq. (\ref{range1}) and in eq.
(\ref{range2}). The only difference between the two cases is that, when
performing the angular integrals (to compute, say the total energy or the
non-Abelian magnetic flux), whenever in the case corresponding to
eq.(\ref{range1}) the result is $4\pi$, the choice corresponding to eq.
(\ref{range2}) has a similar result replaced by $2\pi(2-\delta)$. This
observation will be useful when considering the flat limit in which $R_{0}$ is
very large: indeed one can also consider the limit in which%
\[
R_{0}\rightarrow\infty\ ,\ \ \left(  2-\delta\right)  \rightarrow0\
\]
in such a way that%
\[
\left(  R_{0}\right)  ^{2}\left(  2-\delta\right)  \rightarrow l^{2}%
\ ,\ \ 0<l^{2}<\infty\ .
\]

Hence, the choice of the metric in eq.(\ref{metric}) allows one to study the
Yang-Mills-Higgs system within a tubular topology. In particular, it is possible to
construct configurations which are, at the same time, spherically symmetric
(which is very convenient from the analytical point of view) and can have the
shape of (both closed and open) cylinders (which is important from the
point of view of concrete applications: see \cite{susy} \cite{Shifman:2012zz} and references therein).

These considerations show that the choice of the metric eq. (\ref{metric}),
both with the complete range of angular coordinates eq. (\ref{range1}) and
with the restricted one eq. (\ref{range2}), besides being interesting from
the geometrical point of view is also significant in relation to more applied models. 

As is well known, the system whose action is eq. (\ref{skyrmaction}) in the case of vanishing
Higgs potential $\lambda=0$ admits a BPS completion with the BPS inequality
saturated by monopoles with mass equal to their magnetic charge, $M = Q_{M}$
\cite{Tong:2005un}. In this case the Higgs and gauge fields are related by a
first order BPS equation
\begin{equation}
\label{bps1}B_{i} = D_{i}\Psi.
\end{equation}
 The non-Abelian magnetic charge reads%
\begin{equation}
\label{hedge5}Q_{M} =\int_{\Sigma_{t}}\sqrt{h}d^{3}xTr\left[B_{i}D^{i}%
\Psi\right]  , \quad B_{i} =-\frac{1}{2}\epsilon_{ijk}F_{jk},
\end{equation}
where $h$ is the determinant of the metric induced on $\Sigma_{t}$ (which are
the $t=const$ hypersurfaces) and $B_{i}$ is the non-Abelian magnetic field.
Furthermore the system admits BPS saturated dyons satisfying
\begin{equation}
\label{bps2}B_{i} = \sin(\alpha)D_{i}\psi, \quad E_{i} = \cos(\alpha
)D_{i}\psi
\end{equation}
where $\alpha$ is a parameter chosen to make the bound as tight as possible
and $E_{i} = F_{0i}$ is the electric field. The BPS saturated dyon mass is
\begin{equation}
M_{dyon} = \sqrt{Q_{M}^{2}+Q_{E}^{2}}\;,
\end{equation}
where $Q_{E} $ is the electric charge.

\section{Magnetic solutions}

The ansatz used for the gauge field $A_{\mu}$ and the Higgs field $\Psi$ reads%
\begin{equation}
A_{\mu}=\left(  k\left(  r\right)  -1\right)  U^{-1}\partial_{\mu}%
U\ ,\ \ \Psi=\psi(r)U\ , \label{hedge1}%
\end{equation}%
\begin{equation}
U=\widehat{n}^{i}t_{i}\ ,\ \ U^{-1}=U\ , \label{standard1}%
\end{equation}%
\begin{equation}
\widehat{n}^{1}=\sin\theta\cos\phi\ ,\ \ \ \widehat{n}^{2}=\sin\theta\sin
\phi\ ,\ \ \ \widehat{n}^{3}=\cos\theta\ , \label{standard2}%
\end{equation}
where the $t^{i}$ are the standard Pauli matrices.

This ansatz coincides with the usual 't Hooft-Polyakov ansatz since the factor
$U^{-1}\partial_{\mu}U$ in eq. (\ref{hedge1}) can also be written in the usual
way using the 't Hooft symbols (due to the fact that the $t_{i}$ satisfy the
Clifford algebra). However, the form in eq. (\ref{hedge1}) is more suitable to
analyze the system on non-trivial backgrounds when there is no notion of
Cartesian coordinates.

The Yang-Mills-Higgs system of equations of motion for the ansatz presented in
eqs. (\ref{hedge1}), (\ref{standard1}) and (\ref{standard2}) in the background
shown in eq. (\ref{metric}) reduces to the following coupled system of
two\textit{ autonomous} equations for the two functions $k\left(  r\right)  $
and $\psi(r)$ (we set $v^{2}=1$ from here throughout the rest of the paper):%
\begin{align}
\psi^{\prime\prime}-2\frac{k^{2}\psi}{\left(  R_{0}\right)  ^{2}}%
+\frac{\lambda^{2}}{2}\psi\left(  1-\psi^{2}\right)   &  =0\ ,\label{hedge2}\\
k^{\prime\prime}+\frac{k\left(  1-k^{2}\right)  }{\left(  R_{0}\right)  ^{2}%
}-k\psi^{2} &  =0\ ,\label{hedge3}%
\end{align}
where prime denotes differentiation with respect to $r$. The energy of the
system reduces to
\begin{equation}\label{energy1}
E=\frac{4\pi}{R_{0}^{2}}\int_{0}^{L}\left[  \frac{1}{2}\left(  1+k^{4}\right)
+k^{2}\left(  R_{0}^{2}\psi^{2}-1\right)  +\frac{\lambda^{2}}{8}R_{0}%
^{4}(1-\psi^{2})^{2}+R_{0}^{2}k^{\prime2}+\frac{1}{2}R_{0}^{4}\psi^{\prime
2}\right]  dr,
\end{equation}
and the non-abelian magnetic charge to
\begin{equation}
Q_{M}=4\pi\int_{0}^{L}\frac{\partial}{\partial r}\left[  \psi\left(
k^{2}-1\right)  \right]  dr.\label{charge}%
\end{equation}
Independently of the chosen ansatz the vacuum at vanishing $\lambda$ corresponds to a vanishing gauge field and the Higgs assuming any constant value. This is well known and is the basis of the construction of BPS saturated monopoles in which the Higgs field tends to a non-vanishing constant asymptotically, thus breaking the gauge symmetry in the vacuum. The situation here is subtly different. As we will shortly show the existence of a disconnected boundary at $r=0$, different from the asymptotic one at $r=L$, is sufficient for the existence of monopole solutions even if the Higgs field vanishes there. In this sense, and opposite to standard cases, the field configuration with vanishing energy preserves the gauge symmetry, this being broken at $r=0$. This is a direct consequence of the topology of our system. An equivalent explanation is to remember that in this case the asymptotic flux integral eq. (\ref{charge}) includes a contribution from the border at $r=0$ and thus one can obtain a non-vanishing magnetic charge even though the fields vanish at $r=L$. When there is no potential for the Higgs field and $\lambda=0$, with this ansatz  the vacuum in eq. (\ref{energy1})
corresponds to having a vanishing gauge field $k=1$ and $\psi=0$. This vacuum
is a solution of the equations (\ref{hedge2})-(\ref{hedge3}) and hence we may
find solutions which tend to the real vacuum at large $r$. The true vacuum of the
system in the case of a non-vanishing potential $\lambda\neq0$ consists in
having a gauge field as pure gauge and a Higgs
field at its vacuum expectation value. This physical vacuum corresponds to $k=-1$ and $\psi=1$ in our ansatz. As can be easily seen this doesn't solve the
monopole equations of motion (\ref{hedge2})-(\ref{hedge3}) and therefore a
solution to these equations can never tend to the real vacuum. This is in
sharp contrast to the flat space monopole solution where one can have
asymptotically vanishing energy solutions even if $\lambda \neq0$, the reason being that the damping
coefficient $1/r^{2}$ is here replaced by the constant $1/R_{0}^{2}$. However,
the minimal energy solutions (which we loosely refer to as vacua) will now
depend on the value of the parameters. In the false vacuum where $k=0$,
$\psi=1$ one has $E/L=\frac{2\pi}{R_{0}^{2}}$, whilst in the case in which the
gauge field is pure gauge $k=-1$, $\psi=0$ one has $E/L=\frac{\lambda^{2}\pi
R_{0}^{2}}{2}$. Then if $R_{0}^{2}$ is large (for a fixed $\lambda^{2}$) the
first vacuum has lowest energy. If instead $R_{0}^{2}$ is small the opposite
case might be true (depending on $\lambda$). Remarkably, in the special case
in which $R_{0}^{2}=\frac{2}{\lambda}$ the two vacua are degenerate. 

It is also interesting to note that in the flat case limit $R_{0}\rightarrow\infty$ the equations of motion (\ref{hedge2})-(\ref{hedge3})
simplify considerably. In the equation for the Higgs profile (eq.
(\ref{hedge2})) the interaction term between the Higgs and the gauge field
drops out and one is left with the usual equation for a one-dimensional kink.
Then, in this limit, the gauge field profile $k$ satisfies a Schr\''odinger-like
equation (Eq. (\ref{hedge2}) with $1/R_{0}=0$) in which the kink arising from
the first equation plays the role of the potential. In the flat limit, the present configurations approach domain-walls of the Yang-Mills-Higgs system.  In this limit, the size of the sections of the tube approaches  infinity, the curvature of the sections approaches zero and the energy-density only depends (due to the hedgehog properties) on the longitudinal $r$ coordinate. Moreover, the energy density approaches a delta function (whose position can be taken to be $r=0$).

The following remarks are in order. Analogously to what happens for the
hedgehog ansatz on flat spaces, the ansatz shown in eqs. (\ref{hedge1}),
(\ref{standard1}) and (\ref{standard2}) for our metric reduces the
Yang-Mills-Higgs system (which in principle is a matrix system of coupled
PDEs) to just two coupled ordinary equations for the profiles of the gauge
potential $k(r)$ and the Higgs $\psi(r)$. Moreover, as in the flat space case,
the energy density does not depend on the angles (since all the angular
dependence disappears when computing the trace over the gauge group). In this
sense, this ansatz is a generalized hedgehog ansatz. Furthermore, unlike
what happens when the metric is flat, the system of equations eqs.
(\ref{hedge2}) and (\ref{hedge3}) is autonomous: namely, no explicit power of
the independent coordinate $r$ appears. The reason is that the components of
the metric in eq. (\ref{metric}) do not depend on $r$. Contrarily, in the
standard case of the flat metric in spherical coordinates, the components
(and, consequently, the determinant of the metric) depend on $r$
explicitly\footnote{It is interesting to compare the usual flat system for the
't Hooft-Polyakov monopoles (see, for instance, eq. (3.13a) and (3.13b) in
\cite{rossi}) with the system of equations eqs. (\ref{hedge2}) and
(\ref{hedge3}). The explicit powers of the radius appearing in the flat case
now become constant coefficients (proportional to $\left(  R_{0}\right)
^{-2}$) and, as a consequence of the choice of metric, all terms containing
derivatives of metric determinant vanish.}. The absence of explicit factors of
$r$ in the resulting equations of motion is what allows us to determine the
generic qualitative behaviour of the Yang-Mills-Higgs system using a simple
analytical tool based on an analogy with a two-dimensional Newtonian problem
within a conservative potential, as shown below.

If one defines the
two-dimensional vector $\overrightarrow{x}$%
\[
\overrightarrow{x}=\left(  x,y\right)  =\left(  \frac{k}{c_{1}},\frac{\psi
}{c_{2}}\right)  \ ,\ \ \ \left(  \frac{c_{1}}{c_{2}}\right)  ^{2}%
=\frac{\left(  R_{0}\right)  ^{2}}{2}\ ,
\]
then eqs. (\ref{hedge2}) and (\ref{hedge3}) can be translated to the dynamics
of a two dimensional classical particle subject to a conservative force:%
\begin{align}
\frac{d^{2}x_{i}}{d\tau^{2}} &  =-\frac{\partial V\left(  x,y\right)
}{\partial x_{i}}\ ,\ \ \tau=r\ ,\ \ x_{1}=x\ ,\ \ \ x_{2}%
=y\ ,\label{newtonian1}\\
V\left(  x,y\right)   &  =\frac{1}{\left(  R_{0}\right)  ^{2}}\left[
\frac{x^{2}}{2}-\left(  c_{1}\right)  ^{2}\frac{x^{4}}{4}\right]
+\frac{\lambda^{2}}{2}\left[  \frac{y^{2}}{2}-\left(  c_{2}\right)  ^{2}%
\frac{y^{4}}{4}\right]  -\frac{2\left(  c_{1}\right)  ^{2}}{\left(
R_{0}\right)  ^{2}}\left(  xy\right)  ^{2}\ .\label{newtonian2}%
\end{align}
where the \textquotedblleft effective time" $\tau$ is simply the coordinate
$r$. In this form, the theory of qualitative analysis of dynamical systems
allows one to determine the general behavior of the solutions by just
analysing the plot of the potential $V\left(  x,y\right)  $ even with a
non-vanishing Higgs coupling. It is worth emphasizing here that when
$\lambda=0$, the physical vacuum of the system (which, in terms of the gauge
and Higgs profiles is at $k=1$, $\psi=0$) in terms of the \textquotedblleft
Newtonian" variables corresponds to (we eliminate $c_{1}$ in favor of $c_{2}$
and $R_{0}$) $x=1,\;y=0$. At this point,
\begin{equation}
\partial_{x}V(x,y)=\frac{1}{R_{0}^{2}}-\frac{c_{2}^{2}}{2},\quad\partial
_{y}V(x,y)=0.\label{case1}%
\end{equation}
If we require this point to be an extremum of the potential then we must set
$c_{2}^{2}=2/R_{0}^{2}$. At this point the potential becomes an unimportant
constant which can be removed by a simple shift. However, when $\lambda\neq0$
the case in which the vacuum corresponds to $k=0$, $\psi=1$ which in our
coordinates means $x=0$, $y=1$ for which
\begin{equation}\label{case2}
\partial_{x}V(x,y)=0,\quad\partial_{y}V(x,y)=\frac{1}{2}\left(  1-c_{2}%
^{2}\right)  \lambda^{2}.
\end{equation}
We see that in this case one can also find a particular case in which both
derivatives of the potential vanish by setting $c_{2}^{2}=1$. Finally for the
case in which $k=1,\psi=0$ we revert back to the case shown in eq. (\ref{case1}).
Therefore, many different situations are possible which depend on the
parameters of the system. Accordingly numerous solutions may exist which
interpolate between maxima and minima of the corresponding potentials. We will
return to discuss this analogy with a Newtonian potential once we present some
numerical solutions to the equations of motion. This analogy brings with it useful practical advantages. First of all, it automatically provides a non-trivial conserved quantity (which is of course the ``Newtonian energy" of the system). Secondly, it allows use of the powerful analytical tools of the theory of dynamical systems. This is usually unavailable in the flat Yang-Mills-Higgs system due to the fact that, as already remarked, the corresponding field equations are not autonomous. This is a great benefit from both numerical and analytical points of view since the theory of dynamical systems allows one to analyse the asymptotic behaviour of the solutions and their stability in a very general fashion (without restrictions on the parameters of the theory such as $\lambda$ and $R_{0}$) \cite{dynamical}. We hope to come back on this point on a future publication. 

\subsection{BPS monopoles}

 When $\lambda=0$, the BPS equations (\ref{bps1}) reduce to
\begin{equation}
\partial_{r}k-k\psi=0,\label{BPSv1}%
\end{equation}%
\begin{equation}
\partial_{r}\psi+\frac{1-k^{2}}{R_{0}^{2}}=0.\label{BPSv2}%
\end{equation}
The general solution of the system reads%

\be
k=\exp u\ ,
\ee
\be
\psi=\partial_{r}u\ ,
\ee

where the function $u(r)$ is the inverse of the following integral%
\be\label{integral}
\int_{u(0)}^{u(r)}\left[  2\left(  I_{0}+\frac{\exp\left(  2z\right)
-z}{R_{0}^{2}}\right)  \right]  ^{-1/2}dz=\pm r\ ,
\ee
where $I_{0}$ is an integration constant. Even if
the analytic form of the present BPS magnetic solution is different from the
usual one on flat space without boundaries, the profile of the gauge field as function of the coordinate $r$ along the tube axis is
quite similar to the usual one (see figure \ref{fig11}). The Higgs field appears to differ from the standard flat case, the reason for this is related to the boundary condition at large $r$, we will discuss this further when dealing with numerical solutions. Therefore, this setup realises topologically non-trivial
three-dimensional BPS objects in which the presence of non-trivial boundaries
in the directions transverse to the tube can be analyzed directly, i.e. one has analytical control over both coupling parameters in the action and the topological parameter defining the background geometry. In
particular, as already remarked, the smaller $R_{0}$, the larger
 the plateau in which the profiles remain in the ``false vacuum" along the
$r$ direction.

\subsubsection{Zero modes}

As is well known the monopoles solution in flat space has four bosonic zero modes corresponding to translations of the monopole centre and a gauge rotation of the remaining unbroken symmetry. In fact one can easily prove that the monopole solutions in this topology possess two bosonic zero modes. We will follow a standard procedure suitably adapted to our curvilinear coordinates (see \cite{Balian:2005joa}). Consider the energy functional with $\lambda=0$,
\be 
E = \int d^3x\sqrt{-g}\left(g^{nm}E_nE_m+D_0\phi D_0\phi+g^{nm}(B_nB_m+D_n\phi D_m\phi)\right).
\ee
As we have shown above, for the BPS monopole one can identify the second terms in the above with the constant mass of the monopole
\be
E = M_m+\int d^3x\sqrt{-g}\left(g^{nm}E_nE_m+D_0\phi D_0\phi\right).
\ee
Then, allowing for time dependence of the fields in the static gauge $A_0=0$ one has
\be
E= M_m + \int d^3x \sqrt{-g} \left(g^{nm}\dot{A_n}\dot{A_m}+\dot{\phi}^2\right).
\ee
We consider allowing for a time dependence over the background monopole field, hence we take
\be
A_n(r,\theta, \psi) \rightarrow A_n(r,\theta, \psi)+g^{kl}a_{nk}(r,\theta,\psi)X_l(t)
\ee
\be
\phi(r,\theta, \psi) \rightarrow \phi(r,\theta, \psi)+g^{kl}\chi_n(r,\theta,\psi)X_l(t)
\ee
where the zero modes $a_{nk}$ and $\chi_n$ satisfy the background gauge condition 
\be
D_n a_{kl}^{a}-\epsilon_{abc}\phi^b\chi^c_l=0
\ee
for which one has the standard solution $a_{kl}=F_{kl}$ and $\chi_l = D_l\phi$. Upon substitution inside the energy functional we obtain
\be
E =M_m+ \frac{M_m}{2}\dot{X_l}^2,
\ee
which corresponds to free motion in the three coordinate directions. In our coordinates, these correspond to one translation in the longitudinal $r$ and two in the transverse directions. However, in this topology motion in the trasverse directions is associated to rotations of the solution. These are symmetries of our spherically symmetric solution and thus should not be interpreted as zero modes \footnote{The reduction in number of zero modes is easy to understand since the effective radial coordinate $r$ is here a 1-d coordinate rather than a 3-d one.}.  The final modulus associated to the residual gauge symmetry follows as before, consider the variation
\be
A_n(\vec{x},t) = U(\vec{x},t)A_n(\vec{x})U^{-1}(\vec{x},t)-\frac{i}{e}U(\vec{x},t)\partial_n U^{-1}(\vec{x},t)
\ee
where $\vec{x}=(r,\theta,\psi)$ and
\be
U(\vec{x},t) = \exp\left(ie\gamma(t)\phi\right) \approx 1+ie\dot{\gamma}\phi\delta t.
\ee
In this case $E_n = \dot{\gamma}B_n$ and $D_0\phi=0$ such that
\be
E =M_m+ \frac{1}{2}\dot{\gamma}^2 \int d^3x \sqrt{-g} g^{mn}2B^a_m B^a_n =M_m+ \frac{M_m}{2}\dot{\gamma}^2,
\ee
where in the last line we used the BPS relation for the fields. As a further study, of interest to supersymmetric extensions of this model, it would be desirable to find the fermionic zero modes in the background of this solution. This study is performed by including a Dirac spinor in the fundamental representation of the gauge group with the appropriate Yukawa couplings to the Higgs field. We will delay this study to further work.
\begin{comment}
Therefore, many different situations are possible. For instance, one can have periodic motion, but one cannot have solutions which for asymptotically large Newtonian time $\tau$ approach the vacuum (unless the Higgs field is absent from the very beginning). The reason is that, in order for the two-dimensional Newtonian variable to arrive at the physical vacuum asymptotically (namely, when the Newtonian time $\tau$ approaches infinity) the physical vacuum should be an extremum of the potential in which $\partial_x V$ and $\partial_y V$ vanish simultaneously. On the other hand, there is an open range of parameter space in which $x=0, y=0$ \textbf{this is a trivial solution??} corresponds to a local minimum of the effective potential, to which solutions approach asymptotically. This corresponds to solutions which asymptotically have constant energy density and whose total energy is proportional to the length of the tube itself (as the numerical analysis confirm). In the figures below, there two examples of possible form of the potential. 
\end{comment}

\subsection{Numerical solutions}

\begin{figure}[ptb]
\begin{subfigure}{.5\textwidth}
\centering
\includegraphics[width=0.8\linewidth]{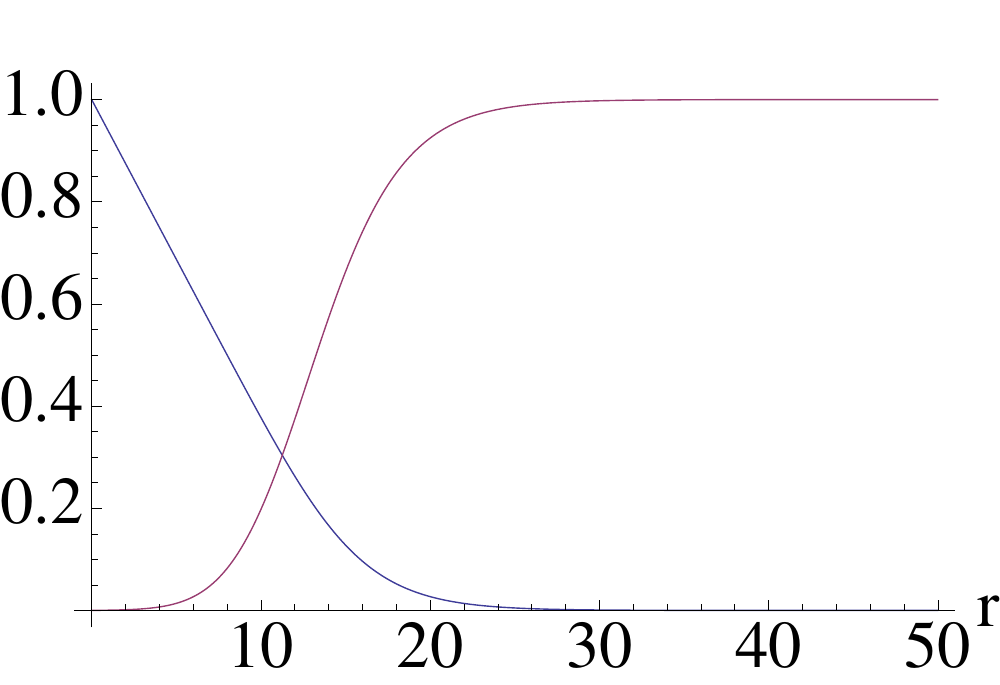}
\end{subfigure}
\begin{subfigure}{.5\textwidth}
\centering
\includegraphics[width=0.8\linewidth]{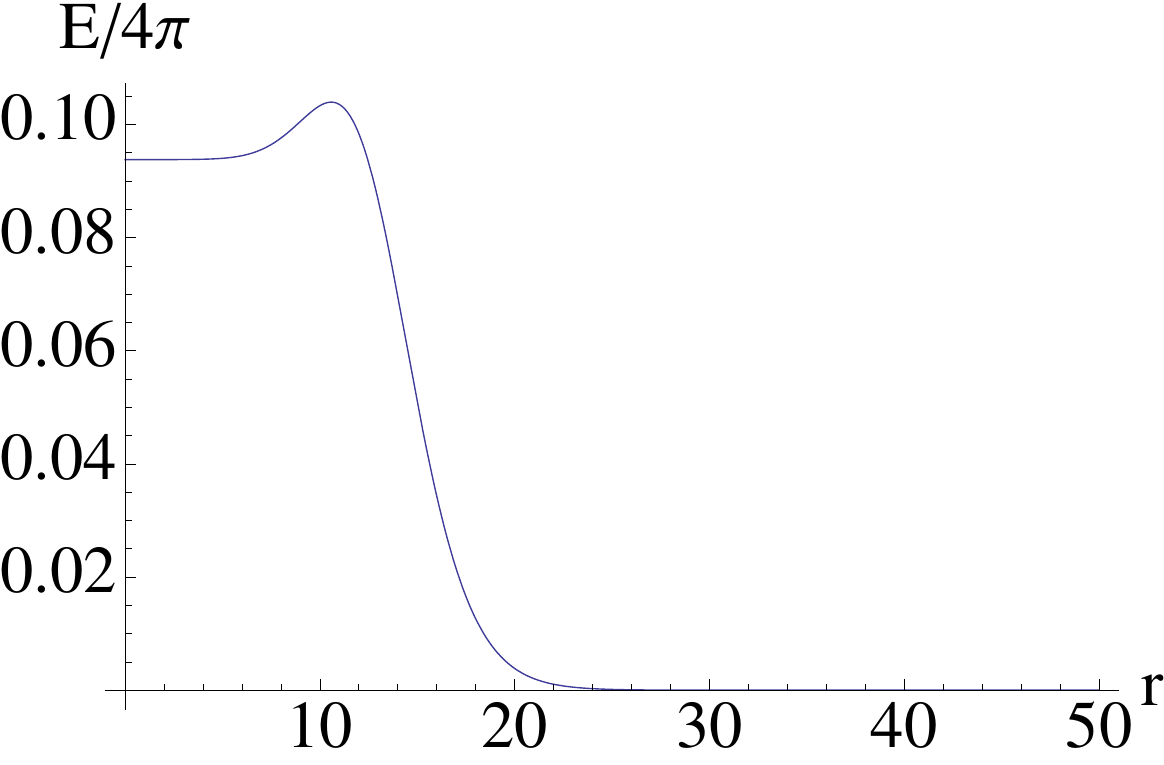}
\end{subfigure}
\caption{Field profiles for a solution with a non-vanishing magnetic charge
and corresponding energy density. The red line corresponds to $k(r)$ whilst
the blue line to the Higgs field $\psi(r)$. The plots correspond to the
numerical values (in appropriate units of $v$) $R_{0} = 4$, $\lambda= 0$,
$L=50$. }%
\label{fig11}%
\end{figure}

Let us proceed to solve the coupled equations (\ref{hedge2})-(\ref{hedge3})
numerically. The numerical solver used throughout the paper is a standard Newton method on a regular grid. We seek solutions which have boundary conditions at $r =0, L$
such that the magnetic charge is not vanishing and which tend to the lowest
energies at $L$ (in the case of $\lambda=0$ this coincides with the vacuum).
Let us begin by the simpler case of vanishing potential $\lambda=0$, then we
seek solutions where
\begin{equation}
\psi(0) =1 , \quad\psi(L) =0
\end{equation}
\begin{equation}
k(0)=0, \quad k(L)=1,
\end{equation}
for which solutions are shown in figure \ref{fig11}. Such boundary conditions (which at first glance could appear unusual) are simply related to the fact that the spatial manifold, in the present case, has two disconnected boundaries. This solution corresponds to a magnetic monopole of charge $|Q|=4\pi$ with an extended core around
$r=0$. The solution is BPS saturated with mass $E/4\pi= |Q|/4\pi= 1$. Making
$R_{0}$ smaller increases the size of the core, thus increasing the region in
which the energy is constant close to $r=0$. This behaviour paints the
following picture, if one assumes that the monopole with unit charge has
constant energy density then upon constricting the monopole to a tubular
geometry this energy density has to fit inside the tube. Once $R_{0}$ shrinks the energy
density has to smear out inside the tube starting from $r=0$ which causes the
solution to possess a large plateau. On the other hand, as already remarked, when $R_{0}$ is very large the configurations approach domain walls.

\begin{figure}[ptb]
\begin{subfigure}{.5\textwidth}
\centering
\includegraphics[width=0.8\linewidth]{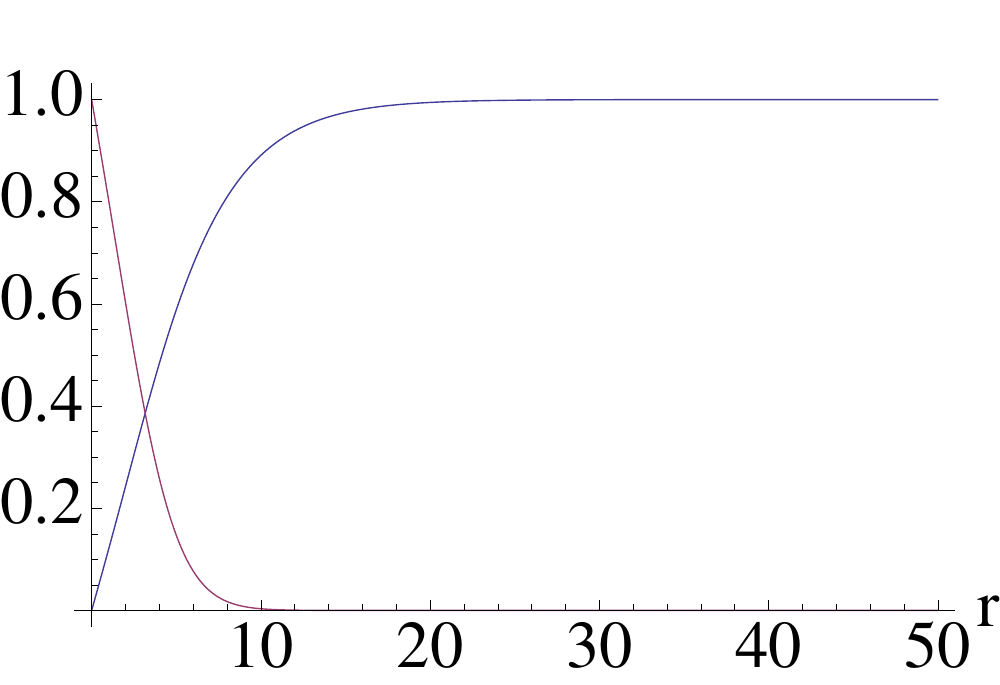}
\end{subfigure}
\begin{subfigure}{.5\textwidth}
\centering
\includegraphics[width=0.8\linewidth]{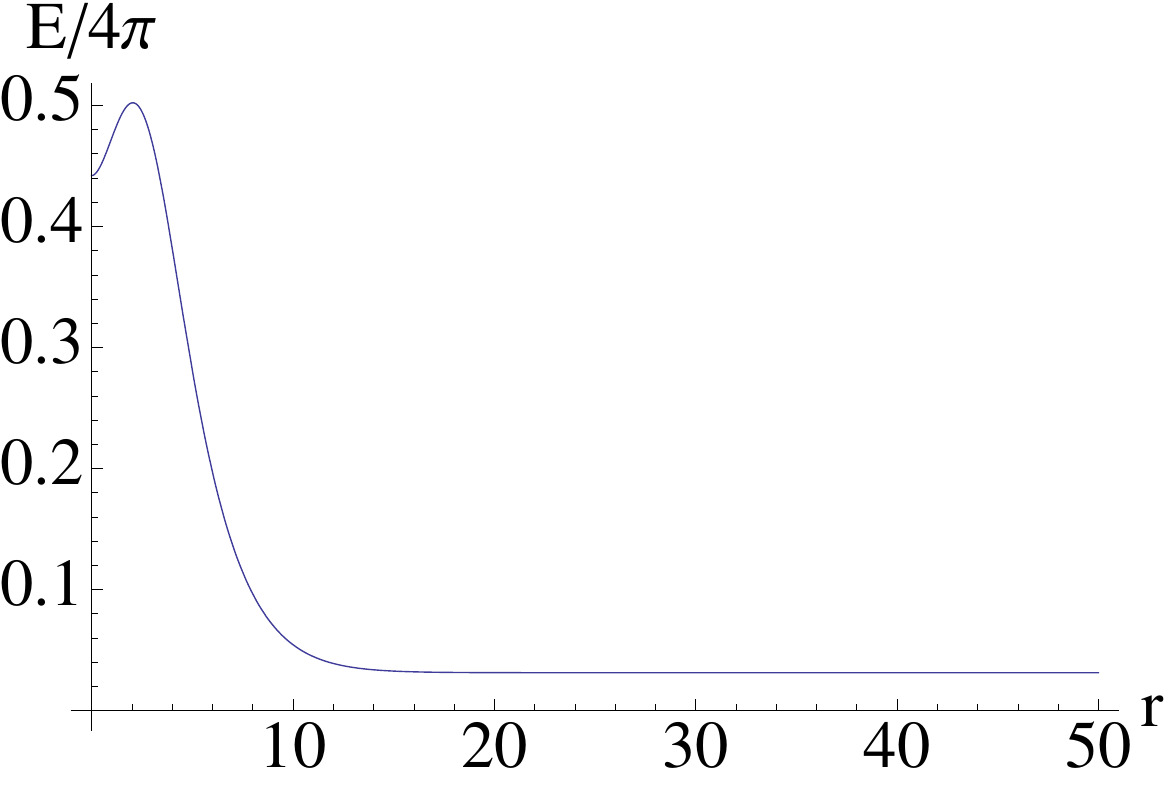}
\end{subfigure}
\caption{Field profiles for a solution with a non-vanishing magnetic charge
and corresponding energy density. The red line corresponds to $k(r)$ whilst
the blue line to the Higgs field $\psi(r)$. The plots correspond to the
numerical values (in appropriate units of $v$) $R_{0} = 4$, $\lambda= 0.3$,
$L=50$. }%
\label{fig2}%
\end{figure}

\begin{figure}[ptb]
\begin{subfigure}{.5\textwidth}
\centering
\includegraphics[width=0.8\linewidth]{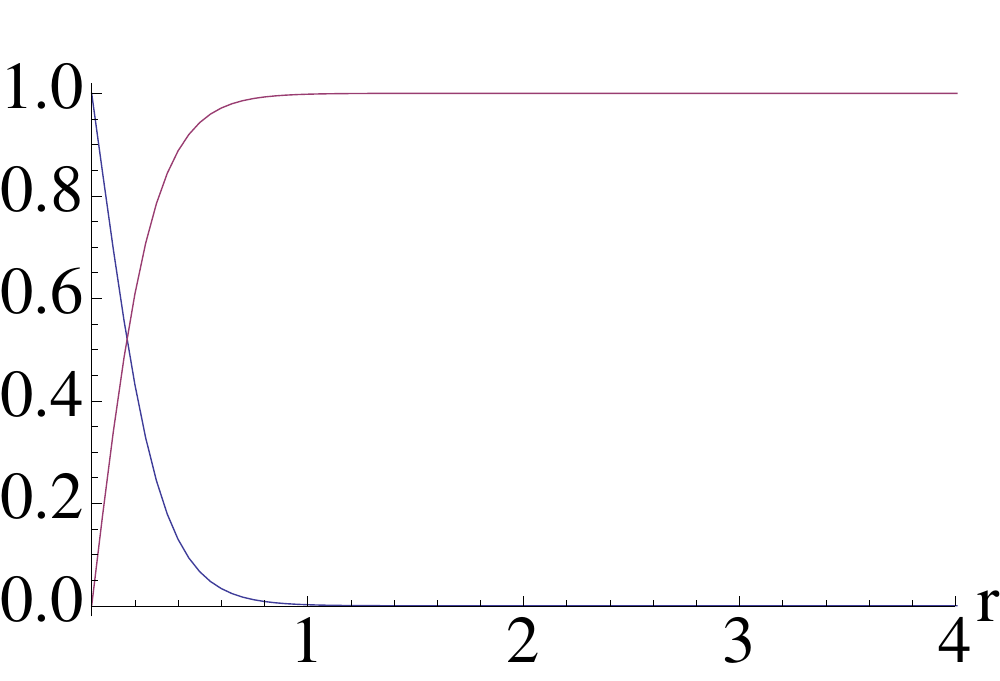}
\end{subfigure}
\begin{subfigure}{.5\textwidth}
\centering
\includegraphics[width=0.8\linewidth]{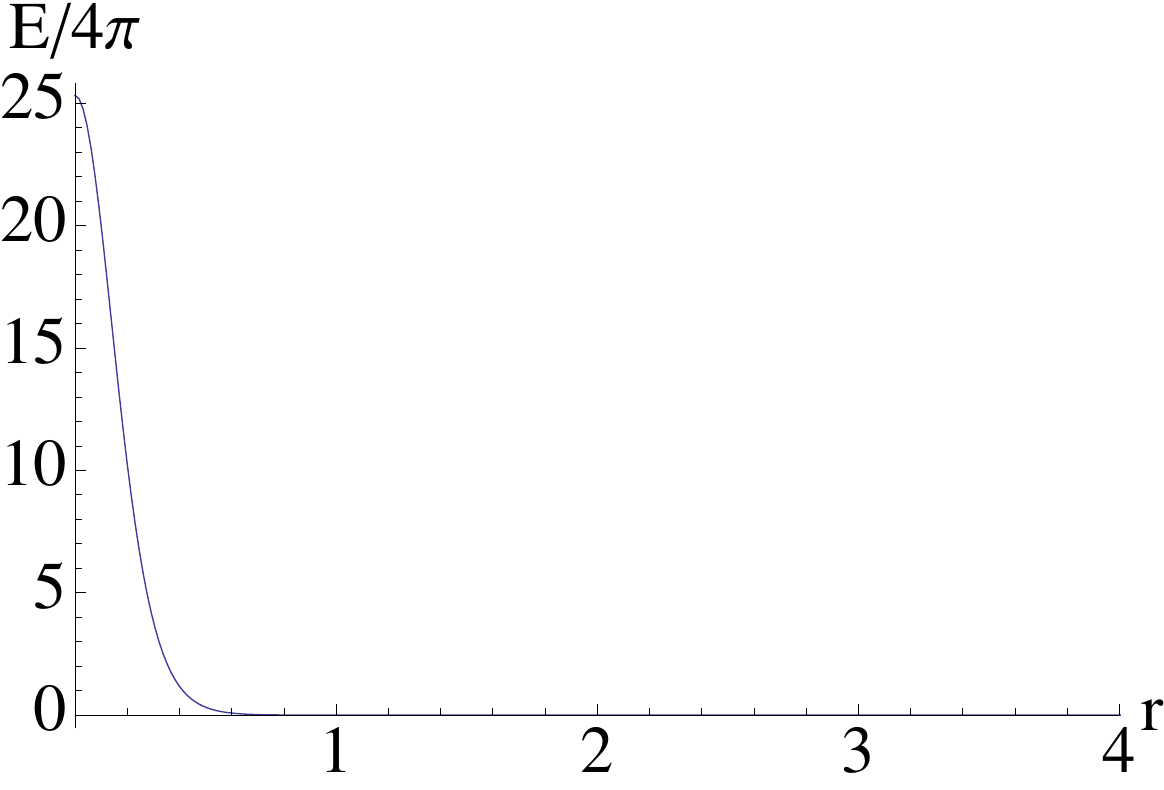}
\end{subfigure}
\caption{Field profiles for a solution with a non-vanishing magnetic charge
and corresponding energy density. The red line corresponds to $k(r)$ whilst
the blue line to the Higgs field $\psi(r)$. The plots correspond to the
numerical values (in appropriate units of $v$) $R_{0} = 0.2$, $\lambda= 0.5$,
$L=50$. The energy tends to a small non-zero constant.}%
\label{fig3}%
\end{figure}

\begin{figure}[ptb]
\begin{subfigure}{.5\textwidth}
\centering
\includegraphics[width=0.8\linewidth]{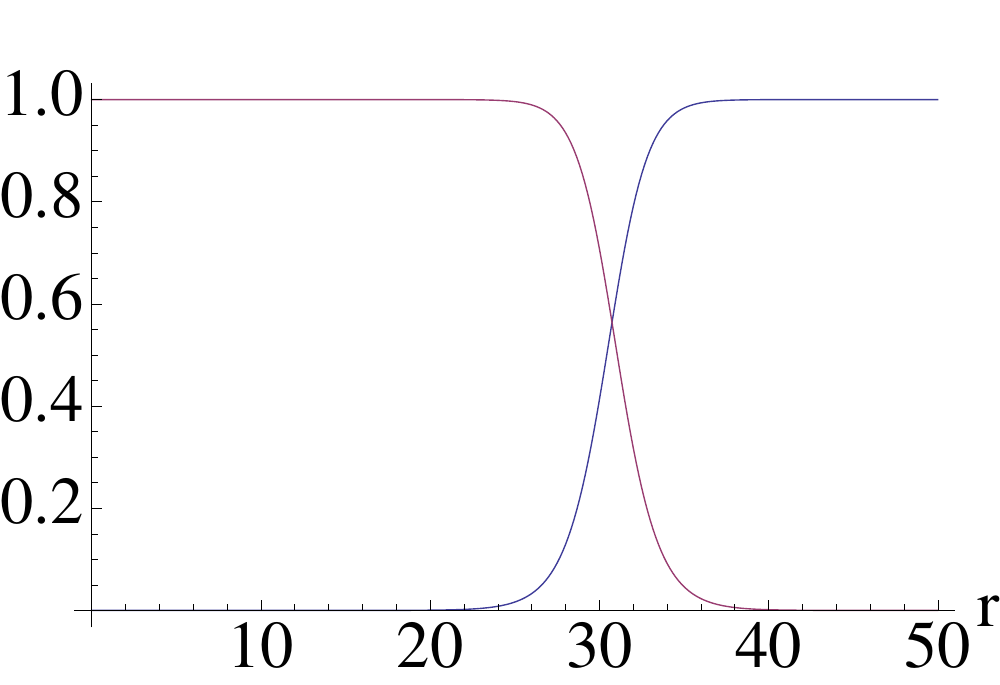}
\end{subfigure}
\begin{subfigure}{.5\textwidth}
\includegraphics[width=0.8\linewidth]{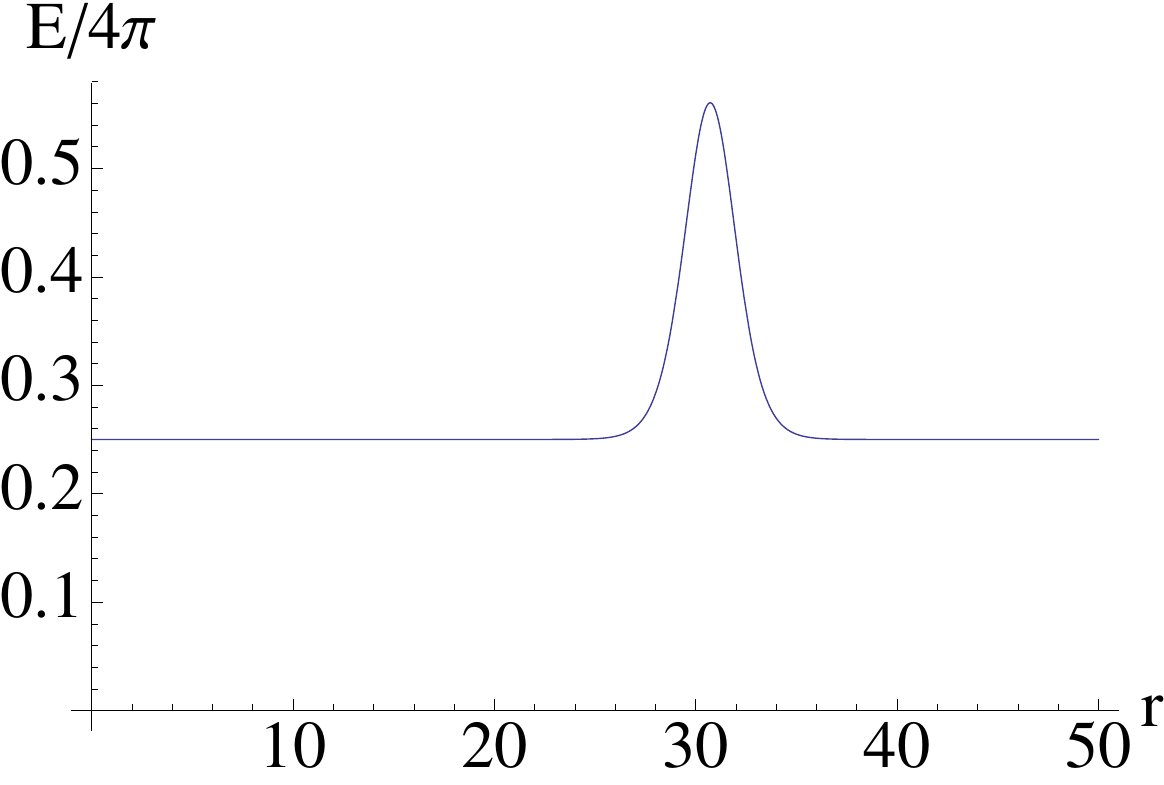}
\end{subfigure}
\caption{Field profiles for a solution with a non-vanishing magnetic charge
and corresponding energy density. The red line corresponds to $k(r)$ whilst
the blue line to the Higgs field $\psi(r)$. The plots correspond to the
numerical values (in appropriate units of $v$) $R_{0} = \sqrt{2}/\lambda$,
$\lambda= 1$, $L=50$. }%
\label{fig4}%
\end{figure}

\begin{figure}[ptb]
\begin{subfigure}{.5\textwidth}
\centering
\includegraphics[width=0.8\linewidth]{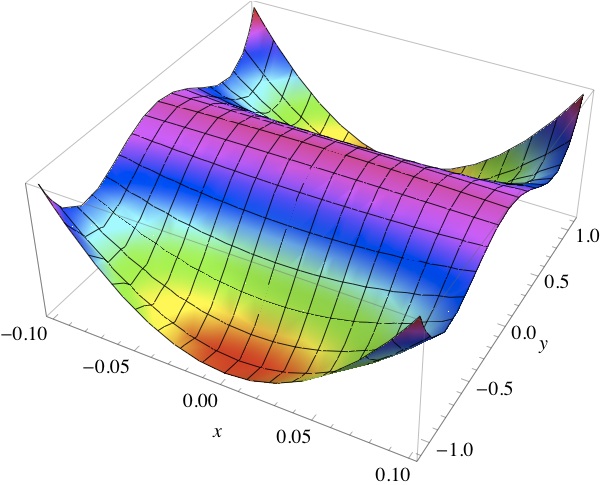}
\label{fig1}
\end{subfigure}
\begin{subfigure}{.5\textwidth}
\centering
\includegraphics[width=0.8\linewidth]{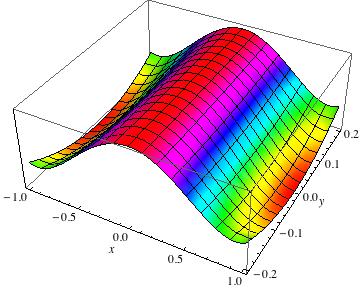}
\label{fig1b}
\end{subfigure}
\caption{Potential profiles corresponding to the parameters chosen for
solutions in figures 2 and 3 respectively. The darker red regions correspond
to minima of the potential. In both cases the solutions can be traced to
trajectories in the Newtonian potential. }%
\label{fig5}%
\end{figure}

\begin{figure}[ptb]
\centering
\includegraphics[width=0.5\linewidth]{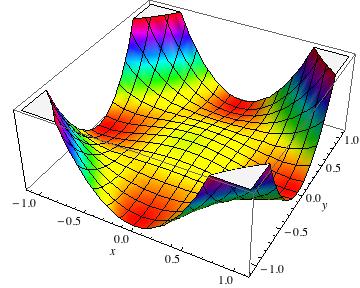}
\caption{Potential profiles corresponding to the parameters chosen for the
solution shown in figure 4. The darker red regions correspond to minima of the
potential. In this special point the solution interpolates between two minima
of the effective potential. }%
\label{fig6}%
\end{figure}

Now let us solve for the case in which $\lambda\neq0$, then we choose
\begin{equation}
\psi(0) =0 , \quad\psi(L) =1
\end{equation}
\begin{equation}
k(0)=1, \quad k(L)=0,
\end{equation}
for which the magnetic charge $|Q| = 1$. Solutions are shown in figures
\ref{fig2} and \ref{fig3}. Here we see that as we expected the energy density
tends to a constant at large $r$ which indicates that the solutions need to be
regulated in the IR. There is also a well defined core radius close to $r=0$.
Our set up has a natural regulator in a finite size tube $L$. Interestingly we
see that there is a large region where the energy density grows linearly with
$L$. Raising $\lambda$, thus raising the Higgs mass narrows the core of the
solution. Conversely, making the cross-section of the tube smaller by
decreasing $R_{0}$ increases the size of the core. In figure \ref{fig3} we
show the solution corresponding to picking boundary conditions such as to pick
out the other vacuum at small $R_{0}$. In this case we find a much narrower
core and a correspondingly higher energy density there. Figure 4 shows a different solution to the same equations in which we set
$R_{0} = \sqrt{2}/\lambda$ with $\lambda=1$, this point is of special
importance as for this choice of $\lambda$ the two vacua in which
$k=0,\;\psi=1$ and $k=1,\; \psi=0$ are degenerate. As shown in the
next section this point is of further importance in the context of solving
equations (\ref{hedge2})-(\ref{hedge3}) analytically. Once again, the solution
requires a IR regulator. As expected these solutions can be traced to
trajectories in the Newtonian potential. Figure \ref{fig5}
shows\footnote{ The effective Newtonian potential defined above is unbounded from below: the obvious reason is that the ``time" of the Newtonian analogy corresponds, in actual fact, to a space-like coordinate} $-V(x,y)$ for both the
parameters used in the solutions shown in figures \ref{fig2} and \ref{fig3},
we see that these solutions correspond to trajectories in which, in the first
case, one starts at the maximum of the potential and tends towards the minimum
(dark red area). Figure \ref{fig6} shows the potential which represents the
solution shown in figure \ref{fig4} interpolating between two degenerate minima. These solutions correspond to the parameters $c_1$ and $c_2$ found using equations (\ref{case1}) and (\ref{case2}).

\begin{comment}These solutions are reminiscent of the standard flux tube solutions (Abrikosov-Nielsen Olesen (ANO) strings) of the flat four dimensional abelian Higgs model. In our case, and in sharp contrast to the (ANO) string, the field $k(r)$ which plays the role of the photon field does not have a vanishing derivative at the origin. However, given the peculiar form of the geometry we are using, this does not lead to a diverging energy as the kinetic terms for $k(r)$ now are not divided by $r$ as would happen in the usual case. The solution therefore represents a magnetic monopole centred at $r =0$ and is independent on the $L$, the size of the cylindrical tube with spherical cross-sections. The energy profile is, as per the standard vortex solution, concentrated around $r=0$. \end{comment}

\subsection{The special point}

Remarkably enough, for a particular choice of $R_{0}$ the system of eqs
(\ref{hedge2}) and (\ref{hedge3}) admits (in the case of non-vanishing Higgs
coupling) a direct analytic treatment in terms of elliptic integrals. Let us
consider the case in which%
\begin{equation}
R_{0}=\frac{\sqrt{2}}{\lambda}\ . \label{special1}%
\end{equation}
In this case, with the ansatz
\begin{equation}
\psi=bk\ , \label{special2}%
\end{equation}
where%
\begin{equation}
b^{2}=\frac{\lambda^{2}}{\lambda^{2}-2} \ ,\ \ \frac{\lambda^{2}}{2}>1\ ,
\label{special3}%
\end{equation}
the system of equations (\ref{hedge2}) and (\ref{hedge3}) reduces to the
single differential equation%
\begin{equation}\label{kink1}
k^{\prime\prime}-k^{3}\left(  b^{2}+\frac{\lambda^{2}}{2}\right)
+\frac{\lambda^{2}}{2}k =0,
\end{equation}
which may be integrated to give
\begin{equation}\label{kink2}
I+\frac{k^{4}}{4}\left(  b^{2}+\frac{\lambda^{2}}{2}\right)  -\frac
{\lambda^{2}}{4}k^{2} =\frac{\left(  k^{\prime}\right)  ^{2}}{2}\ ,
\end{equation}
where $I$ is an integration constant. The general form of equation (\ref{kink1}) admits kink solutions. However these solutions cannot be interpreted as magnetic monopoles because their magnetic charge vanishes. Equation (\ref{kink2}) can be further integrated
directly in terms of an elliptic function%
\begin{equation}
\int_{k(0)}^{k(r)}\frac{dk}{\sqrt{2\left[  I+\frac{k^{4}}{4}\left(
b^{2}+\frac{\lambda^{2}}{2}\right)  -\frac{\lambda^{2}}{4}k^{2}\right]  }%
}=r\ , \label{special5}%
\end{equation}
which fixes the integration constant $I$ by the following relation:%
\begin{equation}
\int_{k(0)}^{k(L)}\frac{dk}{\sqrt{2\left[  I+\frac{k^{4}}{4}\left(
b^{2}+\frac{\lambda^{2}}{2}\right)  -\frac{\lambda^{2}}{4}k^{2}\right]  }%
}=L\ . \label{special6}%
\end{equation}
The above equation determines how the integration constant $I$ depends on the
length of the tube $L$ once the boundary conditions on $k$ are specified. In
particular, solutions corresponding to very large longitudinal length (when, formally, $L$
approaches infinity) are found by choosing the integration constant $I$ in
such a way that $k(L)$ is a double pole of the denominator in eq.
(\ref{special6}). One can also pick the integration constant in order to complete the square inside the denominator. Then for
\be
I = \frac{\lambda^4}{16(b^2+\lambda^2/2)}
\ee
the integral in eq. (\ref{special5}), once inverted, gives 
\be
k(r)= a \tanh\left(-\frac{\lambda(r+C)}{2\sqrt{2}}\right),
\ee
where $C$ is another integration constant which one recognizes as the usual constant determining the kink centre position for this solution and $a=\sqrt{\frac{\lambda^2}{2(b^2+\lambda^2/2)}}$. At this special point and using the ansatz in eq. (\ref{special2}) the energy reduces to
\be
E = 2\pi\lambda^2\int_0^L\left[\frac{(\lambda^2+1)}{2\lambda^2}-\left(\frac{\lambda^2-1}{\lambda^2-2}\right)k^2+\frac{1}{2}\left(\frac{\lambda^4+\lambda^2-4}{(\lambda^2-2)^2}\right)k^4+\frac{2}{\lambda^2}\left(\frac{\lambda^2-1}{\lambda^2-2}\right)k'^2\right]dr,
\ee
which admits a standard completion in terms of a boundary term plus a constant. The boundary term is simply related to the boundary values of the kink profile $k(r)$ whilst the constant is responsible for the linear dependence of the energy on the length of the tube $L$, as expected from the arguments presented in the previous section.

%As already explained (and as the numerical
%analysis confirms), these solutions have energies proportional to the length
%of the tube due to the constant term in the energy-density . The choice
%of the boundary conditions $k(0)$ and $k(L)$ is dictated by physical
%motivations. In particular, one can fix the total magnetic charge (in
%(\ref{charge})): then, taking into account Eqs. (\ref{special2}) and
%(\ref{special3}), one has $I$ as a function of the length of the tube $L$ and
%of the non-Abelian magnetic charge. In this way, for a given charge, one can
%determine the energy as a function of the length of the tube.

\subsubsection{Long-Lived breathers}

One of the most interesting phenomena which occurs in soliton theory in $1+1$
dimensions is the existence of long-lived breathers (two detailed reviews are
\cite{breather1} \cite{breather1.5}). Breathers are long lived oscillating
lumps which lose energy by emitting small amplitude waves. For small enough
amplitudes, the energy emission rate is so slow that it cannot be detected by
a numerical computation\footnote{In the sine-Gordon case, such breathers are
actually exact solutions due to the integrability of the model in $1+1$
dimensions. However, we are interested here in the case of a scalar field with
cuartic potential.}. This intriguing phenomenon at first glance appears to be
quite specific to $1+1$ dimensions since it is known that, in $3+1$ dimensions
breathers have short lifetimes. Hence, one could think that in the case of the
hedgehog sector of the Yang-Mills-Higgs system (which is intrinsically $3+1$
dimensional due to the hedgehog ansatz) one should not be able to find
long-lived breathers. In fact, another benefit of the present geometry is that it allows one to
find such long-lives breathers, as we will shortly show.

To see this, let us generalize the ansatz eq.(\ref{hedge1}) to include
time-dependence in the form
\begin{equation}
A_{\mu}=\left(  k\left(  t,r\right)  -1\right)  U^{-1}\partial_{\mu
}U\ ,\ \ \Psi=\psi(t,r)U\ ,\label{time1}%
\end{equation}
where $U$ is defined as before in eqs. (\ref{standard1}) and (\ref{standard2}%
). In this case, the system's equations of motion in the background metric of
eq. (\ref{metric}) reduce to%
\begin{align}
\left(  \frac{\partial^{2}}{\partial r^{2}}-\frac{\partial^{2}}{\partial
t^{2}}\right)  \psi-2\frac{k^{2}\psi}{\left(  R_{0}\right)  ^{2}}%
+\frac{\lambda^{2}}{2}\psi\left(  1-\psi^{2}\right)   &  =0\ ,\label{time2}\\
\left(  \frac{\partial^{2}}{\partial r^{2}}-\frac{\partial^{2}}{\partial
t^{2}}\right)  k+\frac{k\left(  1-k^{2}\right)  }{\left(  R_{0}\right)  ^{2}%
}-k\psi^{2} &  =0\ .\label{time3}%
\end{align}
Interestingly enough, for the particular case in which $R_{0}$ is chosen as in
eq. (\ref{special1}) the above system reduces to the following single scalar
PDE:%
\begin{equation}
\left(  \frac{\partial^{2}}{\partial r^{2}}-\frac{\partial^{2}}{\partial
t^{2}}\right)  k-k^{3}\left(  b^{2}+\frac{\lambda^{2}}{2}\right)
+\frac{\lambda^{2}}{2}k=0\ .\label{time4}%
\end{equation}
The key observation is that eq. (\ref{time4}) (together with the corresponding
expression for the energy-density) coincides with the equation for a field
theory in $1+1$ dimensions with cuartic interactions (along with the
energy-density), which is known to have long-lived breather solutions
\cite{breather1} \cite{breather1.5}. Hence, the Yang-Mills-Higgs system in the
geometry described by the metric in eq. (\ref{metric}) admits
breather solutions which remain in an oscillatory state for a long time, with
minimal emission of radiation (for a detailed numerical analysis see
\cite{breather2}).

\section{Dyonic solutions}

The present framework allows the inclusion of static dyonic solutions. Let us
consider the following ansatz for the gauge and Higgs fields%
\begin{equation}
A_{0}=\sqrt{2}F(r)U,\ \ A_{i}=\left(  k\left(  r\right)  -1\right)
U^{-1}\partial_{i}U\ ,\ \ \Psi=\psi(r)U\ ,\label{dyons1}%
\end{equation}
where $U$ is defined in eqs. (\ref{standard1}) and (\ref{standard2}) and
$i=1,2,3$ and the factor of $\sqrt{2}$ in $A_{0}$ is added for notational
convenience. The non-abelian magnetic and electric charges are
\begin{equation}
Q_{M}=4\pi\int_{0}^{L}\frac{\partial}{\partial r}\left[  \psi\left(
k^{2}-1\right)  \right]  dr,\label{charge2}%
\end{equation}%
\begin{equation}
Q_{E}=4\pi R_{0}^{2}\int_{0}^{L}\frac{\partial}{\partial r}\left[  \psi
F^{\prime}\right]  dr.\label{charge3}%
\end{equation}

In the tube-shaped domain described by eq. (\ref{metric}) the Yang-Mills-Higgs
equations of motion reduce to%
\begin{align}
\psi^{\prime\prime} &  =2\frac{k^{2}\psi}{\left(  R_{0}\right)  ^{2}}%
-\frac{\lambda^{2}}{2}\psi\left(  1-\psi^{2}\right)  \ ,\label{dyons2}\\
k^{\prime\prime} &  =-\frac{k\left(  1-k^{2}\right)  }{\left(  R_{0}\right)
^{2}}+k\psi^{2}-kF^{2}\ ,\label{dyons3}\\
F^{\prime\prime} &  =\frac{2Fk^{2}}{\left(  R_{0}\right)  ^{2}}%
\ .\label{dyons4}%
\end{align}
Unfortunately, in the dyonic case, the above system of equations cannot be
mapped into a three-dimensional effective mechanical problem with a
conservative potential\footnote{Indeed, the right hand sides of Eqs.
(\ref{dyons3}) and (\ref{dyons4}) cannot be written as derivatives of some
potential depending on the gauge and Higgs profiles}. However, one again one
can see that the present framework brings a very useful simplication. Namely,
despite the fact that the mechanical analogy is unavailable, still, the system
in eqs. (\ref{dyons2}), (\ref{dyons3}) and (\ref{dyons4}) is autonomous (since
no explicit power of the independent $r$ appears) and so the theory of
dynamical systems can be applied
to study the asymptotic qualitative behavior of the above solutions. 

The energy of the system becomes
\begin{equation}
E = \frac{4\pi}{R_{0}^{2}} \int_{0}^{L}\left[  \frac{1}{2}\left(
1+k^{4}\right)  +k^{2}\left(  R_{0}^{2}(\psi^{2}+F^{2})-1\right)
+\frac{\lambda^{2}}{8}R_{0}^{4}(1-\psi^{2})^{2}+R_{0}^{2}k^{\prime2}+\frac
{1}{2}R_{0}^{4}F^{\prime2}+\frac{1}{2}R_{0}^{4}\psi^{\prime2}\right]  dr.
\end{equation}

Once again the dyonic system is simpler than the flat-space one (see, for
instance, \cite{rossi}) due to the fact that in eqs. (\ref{dyons2}),
(\ref{dyons3}) and (\ref{dyons4}) no explicit power of the radius $r$ appears
(neither, of course, do the angular coordinates thanks to the hedgehog
property). The vacuum structure follows similarly to the purely magnetic case.
Clearly, the real vacuum at $\lambda=0$ will have $F=0$. When $\lambda\neq0$
then the minimal energy density is achieved when $F=c$, where $c$ is a
constant. Equation (\ref{dyons4}) also admits a solution in which
$F(r)$ is linear in $r$ when $k=0$.

\subsection{The BPS solutions}

The BPS equations (\ref{bps2}) corresponding to $\lambda=0$ reduce to
\begin{equation}
\partial_{r}k-\sin(\alpha)k\psi=0,\label{bpsdyon}%
\end{equation}%
\begin{equation}
\sin({\alpha})\partial_{r}\psi+\frac{1-k^{2}}{R_{0}^{2}}=0,
\end{equation}%
\begin{equation}
F=\cos(\alpha)\psi.\label{bpsdyon2}%
\end{equation}
The general solution of the system reads%

\be
k=\exp u\ ,
\ee
\be
\sin(\alpha)\psi=\partial_{r}u\ ,\ \ F=\cos(\alpha)\psi\ ,
\ee
where, as in the purely magnetic case, the function $u(r)$ is given by the inverse of
the integral in eq.(\ref{integral}). 

\subsection{Numerical Solutions}

Let us proceed to solve equations (\ref{dyons2})-(\ref{dyons4}) numerically. A
dyon solution will be a solution which at large $L$ tends to the real vacuum
and has non-vanishing electric and magnetic fields. In the case of vanishing
$\lambda=0$ we can further simplify the equations by using an ansatz where
$F = \psi$, then we seek solutions with boundary conditions
\begin{equation}
\psi(0)=1, \quad\psi(L) = 0
\end{equation}
\begin{equation}
k(0)=0, \quad k(L)=1.
\end{equation}
Figure \ref{fig3} shows such a solution corresponding to a well localized
dyon. This solution is valid for any $L$ extending out to infinity. The effect
of decreasing $R_{0}$ is to make the core size of the dyon smaller. Then the
BPS charges are
\begin{equation}
Q_{M} = 4\pi, \quad Q_{E} \approx-4\pi(1.232)
\end{equation}
and numerically we find that
\begin{equation}
\frac{E_{dyon}}{4\pi} = 1.704 > \frac{1}{4\pi}\sqrt{Q_{M}^{2}+Q_{E}^{2}} =
1.587.
\end{equation}
Therefore we conclude that, even for $\lambda=0$ this solution is not BPS
saturated. This is to be expected, not every solution of equations
(\ref{dyons2})-(\ref{dyons4}) is also a solution of (\ref{bpsdyon}%
)-(\ref{bpsdyon2}) (the inverse statement however is true). Specifically for
our solution $F=\psi$ which corresponds to $\alpha= 0 + 2\pi n $ in equation
(\ref{bpsdyon2}). Then this is only a solution of the full set of BPS
equations if $k=1$ identically, which is clearly not the case of our numerical solution.

Let us switch on a potential and set $\lambda\neq0$. Then we seek solutions
where
\begin{equation}
\psi(0)=0, \quad\psi(L) = 1
\end{equation}
\begin{equation}
k(0)=1, \quad k(L)=0.
\end{equation}
\begin{equation}
F(0)=1, \quad F^{\prime}(L)=0.
\end{equation}
A solution satisfying these boundary conditions is shown in figure \ref{fig8}.
As expected this solution has non-vanishing energy density at large $r$ where
it tends to a constant $E = 4\pi/R_{0}^{2}$. Once again, the energy in this
region will be proportional to the tube length $L$, which serves as the IR
regulator. Increasing $\lambda$ decreases the size of the core whilst
increasing $R_{0}$ makes the asymptotic value of $F \rightarrow\psi$ and
decreases the asymptotic value of the energy whilst increasing its ``core",
see figure \ref{fig9}. As expected, in the flat space limit the energy density
should vanish at infinity.

\begin{figure}[ptb]
\begin{subfigure}{.5\textwidth}
\centering
\includegraphics[width=0.8\linewidth]{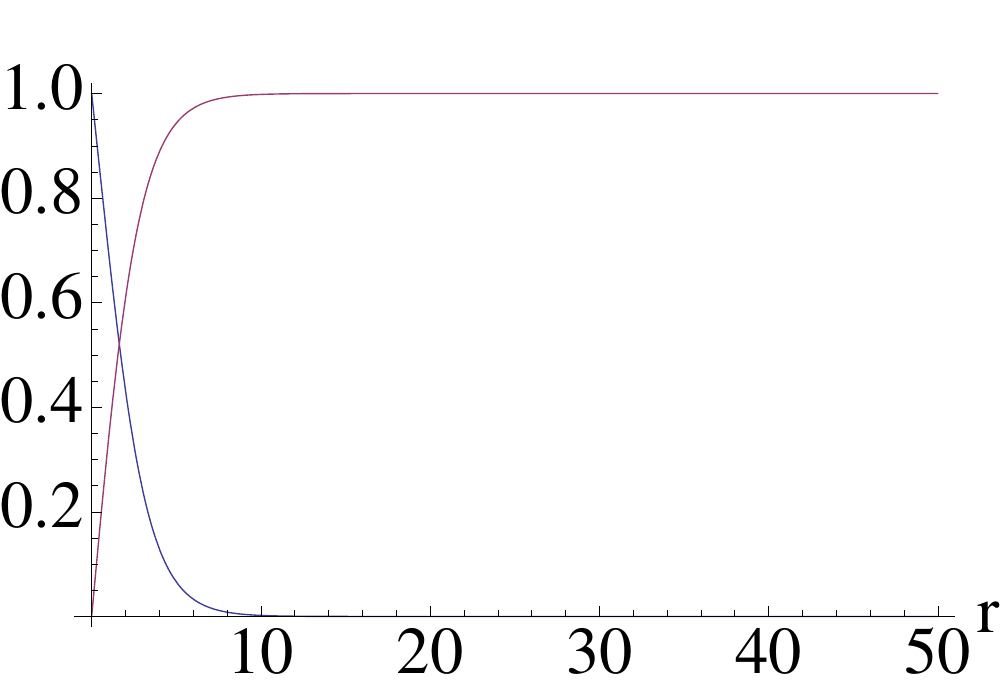}
\end{subfigure}
\begin{subfigure}{.5\textwidth}
\includegraphics[width=0.8\linewidth]{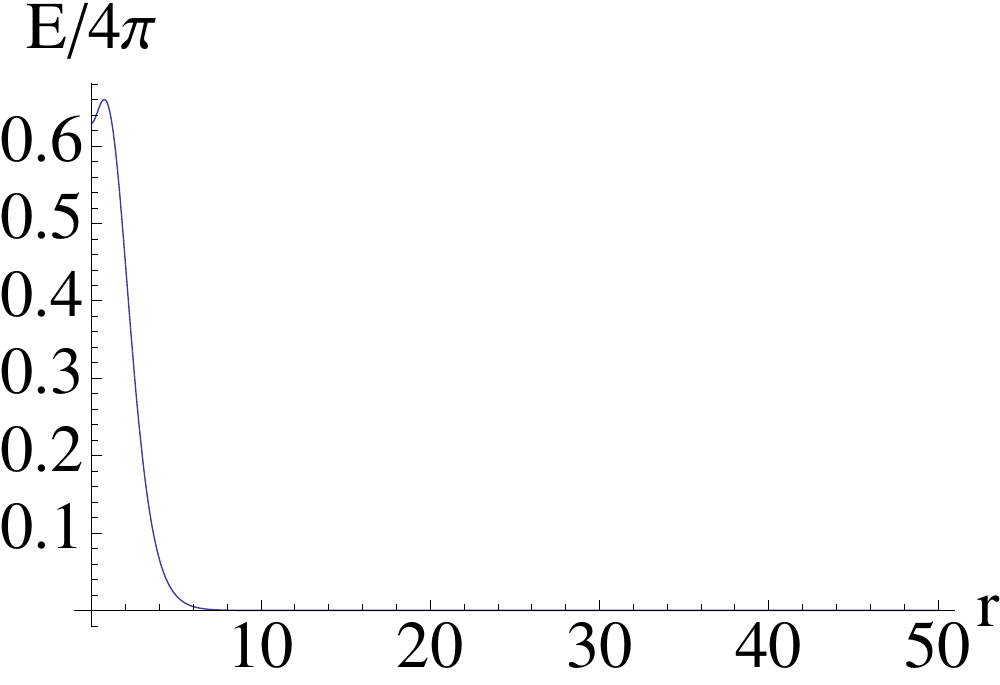}
\end{subfigure}
\caption{Field profiles for dyon solution. The red line corresponds to $k(r)$
whilst the blue line to the Higgs and electric component of the gauge fields
$\psi(r) = F(r)$. The plots correspond to the numerical values (in appropriate
units of $v$) $R_{0} =2$, $\lambda= 0$, $L=50$. }%
\label{fig7}%
\end{figure}

\begin{figure}[ptb]
\begin{subfigure}{.5\textwidth}
\centering
\includegraphics[width=0.8\linewidth]{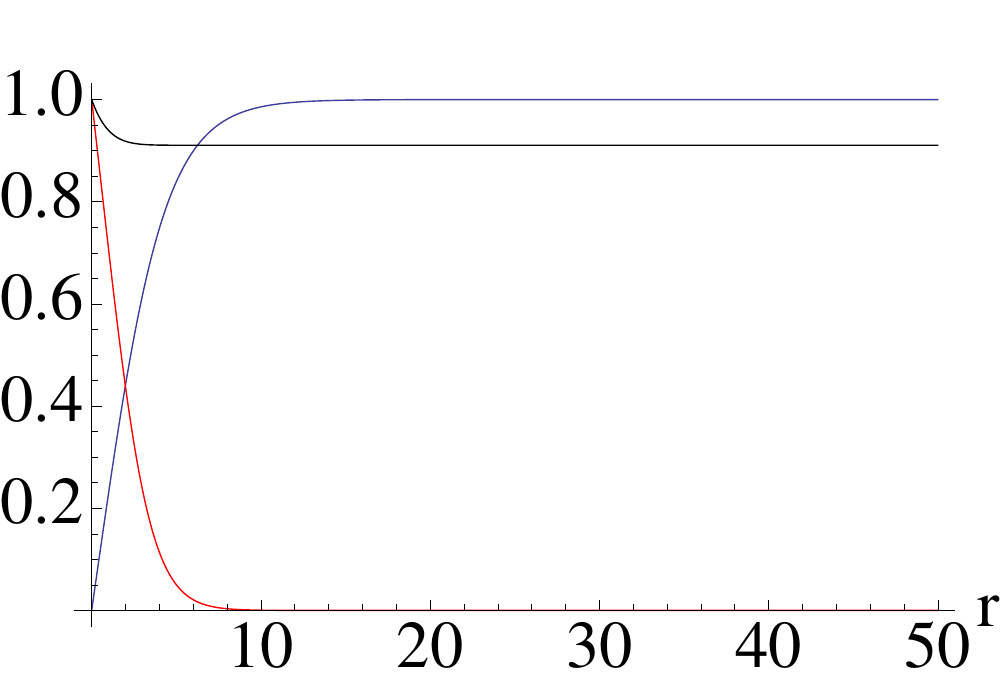}
\end{subfigure}
\begin{subfigure}{.5\textwidth}
\includegraphics[width=0.8\linewidth]{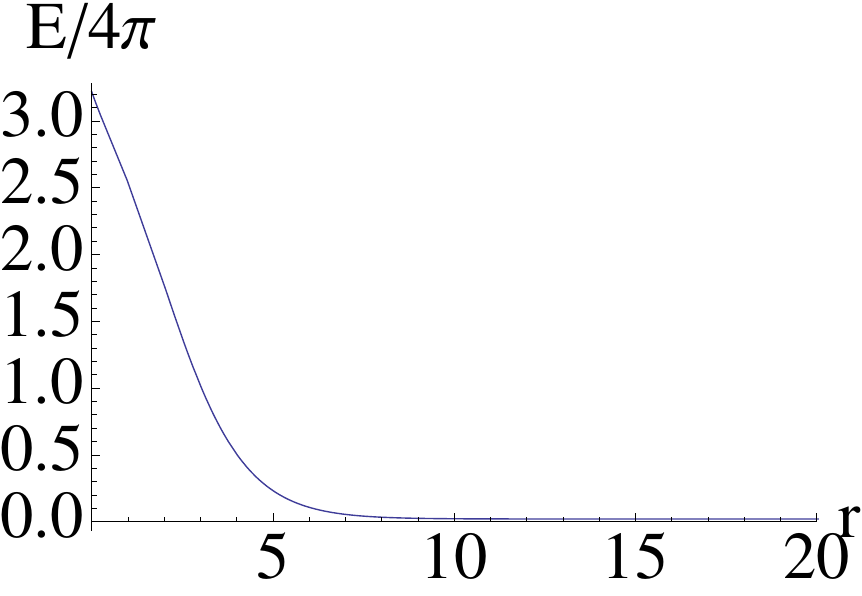}
\end{subfigure}
\caption{Field profiles for non localised dyon solution and corresponding
energy density. The red line corresponds to $k(r)$, the blue line to the Higgs
field $\psi(r)$ and the black like to $F(r)$. The plots correspond to the
numerical values (in appropriate units of $v$) $R_{0} = 5$, $\lambda= 0.5$,
$L=50$. }%
\label{fig8}%
\end{figure}\begin{figure}[ptb]
\begin{subfigure}{.5\textwidth}
\centering
\includegraphics[width=0.8\linewidth]{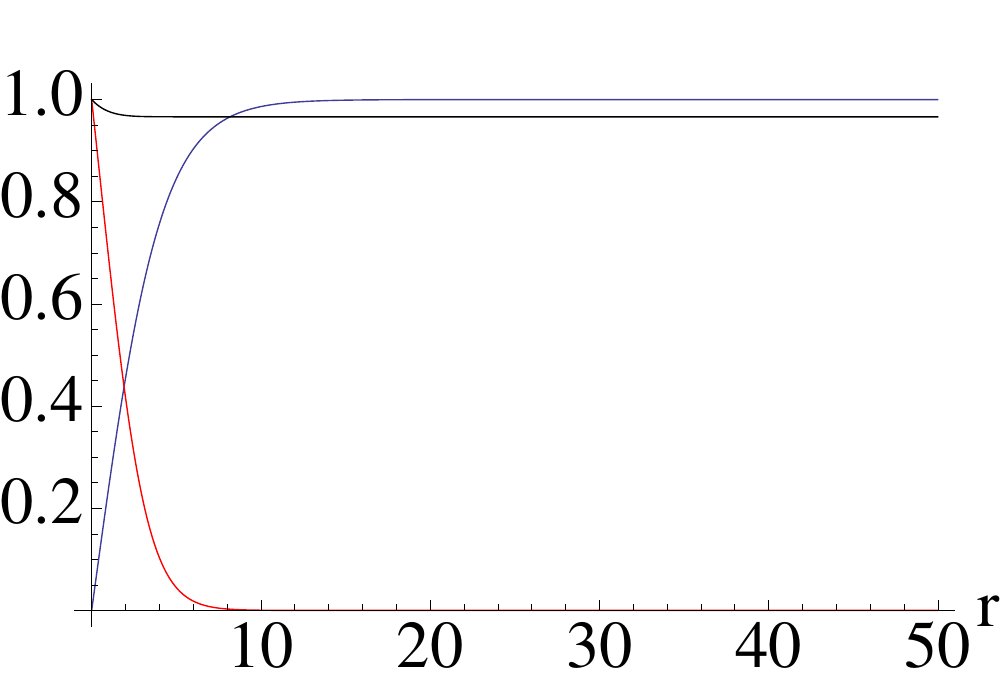}
\end{subfigure}
\begin{subfigure}{.5\textwidth}
\includegraphics[width=0.8\linewidth]{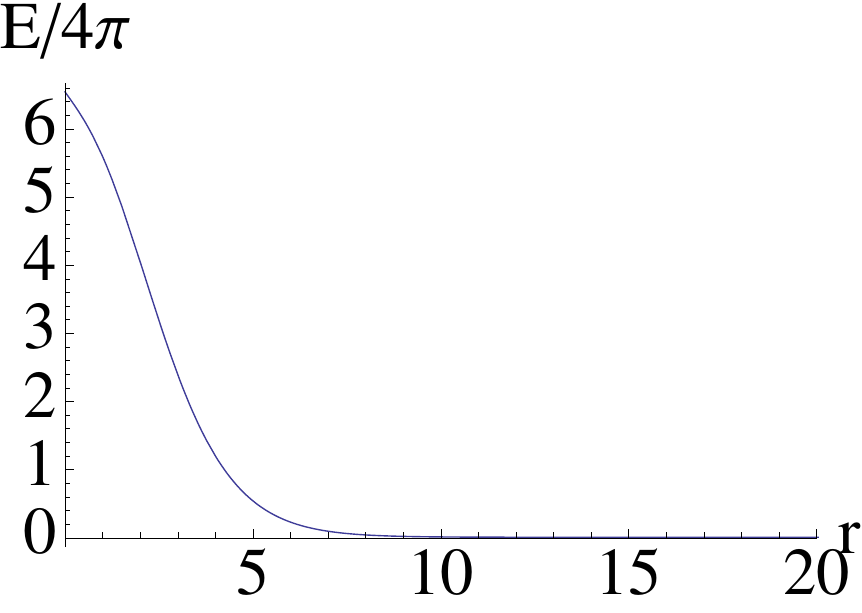}
\end{subfigure}
\caption{Field profiles for non localised dyon solution and corresponding
energy density. The red line corresponds to $k(r)$, the blue line to the Higgs
field $\psi(r)$ and the black like to $F(r)$. The plots correspond to the
numerical values (in appropriate units of $v$) $R_{0} = 8$, $\lambda= 0.5$,
$L=50$. }%
\label{fig9}%
\end{figure}

\section{Conclusions}

We considered exact and numerical solutions of the four-dimensional Yang-Mills-Higgs
system with and without the Higgs coupling $\lambda$ in the background of $\mathbb{R}\times S^2$.
These non-Abelian magnetic configurations can live within both
closed and open tubes whose sections are spherical caps admitting a smooth
flat limit. When $\lambda=0$ the solutions describe magnetic monopoles with a finite energy density. When one switches on a Higgs potential, and unlike what happens in flat case, the energy of the solutions grows linearly with $L$ at large distances, which thus serves as a good cut-off. In the flat limit $R_0\rightarrow \infty$, these configurations approach domain-walls of the Yang-Mills-Higgs system. As $R_0$ shrinks this has the effect of constraining the monopole to the tubular geometry which is seen as a large plateau appearing in its energy density profile.
The dramatic simplification that this geometry provides not only
renders the analytic treatment of the autonomous non-linear coupled system possible but
also simplifies the numerical analysis. The theory of dynamical systems (which, in the standard case, is unavailable) provides many powerful techniques which, for instance, allow study of the asymptotic behaviour as well as the stability of the solutions in a general fashion (namely, without restrictions on the parameters of the theory). When the system involves only a magnetic
component of the gauge field we have shown that the equations can be treated
in the language of a classical particle subject to a conservative force. If
the radius of curvature of the geometry is chosen to satisfy a particular
relation shown in eq.(\ref{special1}) then the system can be further
reduced to a single PDE and its energy can be determined by solving an
elliptic integral without even requiring the explicit dependence on the
profile functions in terms of $r$. At this special point the solutions of the system are kink-like but cannot be interpreted as magnetic monopoles as their non-Abelian charge vanishes. For the magnetic choice of gauge function and for a special choice of the curvature radius,
we have explicitly shown that the hedgehog property admits the inclusion of
time dependence in the field profiles which leads to breather-like solutions
which remain in an oscillatory state for a remarkably long time, with minimal
emission of radiation. Finally, static dyonic configurations with an electric
component can be constructed analytically and numerically and satisfy similar properties. An interesting extension to this work is to consider a similar set-up in the background of $S^1\times S^2$ where one expects to see solutions with periodic boundary conditions. This is achieved simply by identifying the points at $r=0$ and $r=L$, thus one expects similar properties observed in the $\mathbb{R}\times S^2$ system. We defer this investigation to future work.

\subsection*{Acknowledgements}

This work has been funded by the Fondecyt grants 1120352 and 3140122. The
Centro de Estudios Cient\'{\i}ficos (CECs) is funded by the Chilean Government
through the Centers of Excellence Base Financing Program of Conicyt.

\end{document}